\documentclass[twocolumn,nofootinbib,amsmath,amssymb,aps]{revtex4-1}
\usepackage{mathtools,cancel}
\usepackage{amsmath}
\usepackage{slashed}
\usepackage{alltt}
\usepackage{graphicx}
\usepackage{dcolumn}
\usepackage{bm}
\usepackage{color}
\usepackage{mathrsfs}
\usepackage{hyperref}
\usepackage[dvipsnames]{xcolor}
\usepackage[all]{xy}

\hypersetup{
breaklinks=true}

\begin{document}


\title{Large Gauge Symmetries of an Asymptotically de Sitter Horizon:\\ An Extended First Law of Thermodynamics }

\author{Fatemeh Mahdieh}
\email{f\_mahdieh@sbu.ac.ir}
 \affiliation{Department of Physics, Shahid Beheshti University, G.C., Evin, Tehran 1983969411, Iran}

\author{Hossein Shojaie}
\email{h-shojaie@sbu.ac.ir}
\affiliation{Department of Physics, Shahid Beheshti University, G.C., Evin, Tehran 1983969411, Iran}


\begin{abstract}
In this paper, we show that a universe with a dynamical cosmological constant approaching pure de Sitter at timelike infinity, enjoys an infinite dimensional symmetry group at its horizon. This group is larger than the usual $SO(4,1)$ of pure de Sitter. The charges associated with the asymptotic symmetry generators are non-integrable, and we demonstrate that they promote an extended version of the first law of thermodynamics. This contains four pairs of conjugate variables. The pair $(\Theta,\Lambda)$ corresponds to the change in the cosmological constant and its conjugate volume $\Theta$. The contribution of the surface tension of the horizon and its conjugate parameter surface area make a pair $(\sigma, A)$. The usual conjugate variables $ (T, S) $, $ (\Omega, J) $ and a term $ \partial_v \delta S $ corresponding to entropy production, are included. In addition, this extended first law describes the non-conservative behaviour of the asymptotic charges in non-equilibrium.
\end{abstract}

\maketitle

\section{Introduction}\label{Introduction}
The extra symmetries of asymptotically flat spacetimes were first discussed by Bondi, van der Burg, Metzner and Sachs in the early 60's~\cite{Bondi62,Sachs62,Sachs1962}. In fact they were discovered in a ``fortunate" failure of trying to obtain the Poincar\'e group as the exact symmetry group of asymptotically flat spacetimes.

The mathematical machinery for asymptotic symmetries has been studied further and developed extensively by different research groups~\cite{BarnichBrandt,Barnich-BMSalgebra,Barnich-Aspects,WaldZoupas,WaldIyerBHENTROPY,Madler:2016xju,Barnich-STcall,Ashtekar:1996cd,Brown:1986nw,Kapec2014-SR,Andy2014Soft} and references therein. But the physical concept of what these extra symmetries describe has received additional attention after Strominger et.al.~\cite{Andy-book,Kapec2014-SR,Andy2014Soft,Strominger2014-ASofYM,He2015,Sasha2014,He2014Newsym,KapecNew,Pasterski2016,BMSinvGravScat,Kapec2014-SR,HPS1} related three notions from completely different areas together. They introduced a triangle with asymptotic symmetries being one of the corners, and the other two are soft theorems and the memory effect. This area of research has thereafter developed in work done by various groups at null infinity, spatial infinity~\cite{Henneaux2018} and the black hole horizon~\cite{HPS2,Donnay2016,Donnay:2015abr}. The asymptotic limit is usually a boundary in spacetime, and these symmetries are generated by vector fields that are asymptotically Killing on the boundary. In asymptotically flat spacetimes, null infinity or spacelike infinity is taken to be the boundary, and asymptotically Killing vector fields on this boundary are derived. These generators can be written as an expansion of a parameter that approaches zero in the asymptotic region. Accordingly, $\mathcal{L}_\xi g$ is also of order of this parameter, and not exactly zero, but asymptotically vanishes. 

This is the case in asymptotically flat spacetimes at their boundaries with the parameter being $\frac{1}{r}$. This gives $\mathcal{L}_{\xi}g_{\mu\nu} = \mathcal{O}(1/r)$. Likewise, if the horizon of a black hole is taken to be the boundary, the same expansion can be written in terms of a radial parameter $\rho$, that describes the horizon through the equation $\rho = 0$.

In asymptotically flat spacetimes, the BMS group is the semi-direct product of the Poincar\'e group and super translations, i.e. $BMS = Poincar\Acute{e} \times ST $. This group is manifest when studying the symmetries of an asymptotically flat spacetime at null infinity. Here, the supertranslations have been derived by Bondi, Metzner and Sachs~\cite{Sachs62,Bondi62}. This group can additionally be extended to give Virasoro-like generators referred to as superrotations~\cite{Kapec2014-SR}. In the case of spacelike infinity in an asymptotically flat spacetime, it has been shown that in the canonical formalism, by considering suitable matching conditions, supertranslations can be recovered at spacelike  infinity of a dynamical asymptotically flat spacetime\cite{Henneaux2018}. In this context, electromagnetism has also been included~\cite{HTEMSI}. The study of asymptotic symmetries at spatial infinity was worked out covariantly in~\cite{BMS4SI}, where the relation between the symmetries at past and future null infinity was discussed and again electromagnetism was included ~\cite{HTASEMSI}. 

The physical concept of the asymptotic symmetries can be understood as the transition between different asymptotically flat states of an evaporating black hole. In other words, these symmetries characterize the vacuum transitions in the evaporation of a black hole. A boosted radiating black hole~\cite{HPS2} and as shown recently, a rotating black hole~\cite{HPHS}, are other examples of spacetimes possessing these extra symmetries.  

One should mention that, due to the lack of a timelike Killing vector field in the geometry of a dynamical spacetime, there are difficulties in defining the entropy and surface gravity, which result in issues with assigning a temperature to the black hole horizon. There are conserved charges associated with the asymptotic symmetries of dynamical spacetimes, that are given by Noether's theorem. Using the conserved charges, an analogue of the first law of thermodynamics can be obtained~\cite{WaldIyerBHENTROPY,WaldEntropy}.

To highlight our choice of setup for a dynamical system, we note that maximally symmetric spacetimes are of interest to theoretical physicists as well as cosmologists. Regarding the n-dimensional pseudo-Reimannian manifolds, these are Minkowski, de Sitter and anti-de Sitter spacetimes, with zero, positive and negative curvatures respectively. These spacetimes can be seen as the ``ground states" of general relativity~\cite{carroll04}. On the other hand, observations provide that the fate of our Universe is similar to de Sitter spacetime. According to the current values of the density of matter and cosmological constant, it seems that the locally gravitationally bound group of galaxies, i.e. cluster of galaxies, are the typical structures that will remain bound as the Universe expands. Therefore, one can take the Universe to be undergoing a process that eventually tends to pure de Sitter. 

In this work we aim to find the extra symmetries of a dynamical de Sitter-like spacetime at its horizon in order to study thermodynamic aspects of this dynamical system.\footnote{The asymptotic symmetries of de Sitter spacetime have been studied in~\cite{Ferreira16} and references therein. However, in this paper we consider the de Sitter horizon as the asymptotic boundary, whereas the previous work mentioned has mainly studied the asymptotic symmetries at null infinity.} It is worth noting that extra symmetries on a black hole horizon have been studied in ~\cite{Donnay:2015abr,Donnay2016}. Here we consider a different geometry and constraints in order to retrieve pure de Sitter at timelike infinity. Furthermore, we do not dismiss the time dependency of the metric components to ensure we have a dynamical system settling in pure de Sitter.  

In a dynamical spacetime such as Friedmann-Lemaître-Robertson-Walker (FLRW), there is no timelike Killing vector field in general and the surface gravity cannot be defined in the usual manner. However, there are several ways to look at the surface gravity, in which not all methods reproduce the expected temperature in the static limit~\cite{Neilsen07,Visser,Nielsen,cai,Fodor,Helou}. Therefore there is not yet a conclusion to a preferred or correct approach to defining the surface gravity in the dynamical case, and therefore we lack a description for the thermodynamics of such a spacetime.
    
The static patch of pure de-Sitter has a temperature of $2\pi/\kappa$~\cite{HawkingGibbons}, which is the analogue of the black hole temperature. Since all dynamical black holes ultimately settle in a stationary state, in the language of equilibrium and non-equilibrium thermodynamics, a stationary black hole with temperature $T$ can be considered as the analogue of an equilibrium state. Similarly, the static patch of pure de Sitter spacetime with temperature $T$ is also the analogue of an equilibrium state in an inflating universe. To present an analogue of an extended first law of thermodynamics in a de Sitter-like dynamical spacetime, we turn to non-equilibrium thermodynamics. Here, by a de Sitter-like spacetime, we mean a geometry that tends to pure de Sitter at timelike infinity. 

We show that the near-horizon geometry of a de Sitter-like spacetime admits asymptotic symmetries at its horizon. In other words, this spacetime possesses a larger symmetry group than $SO(4,1)$ of pure de Sitter. The conserved charges related to these symmetries can be obtained by means of Noether's theorem and constructing the symplectic form \cite{Banados,WaldZoupas}. In a similar manner, for a non-stationary perturbation to the near-horizon geometry of a de~Sitter-like spacetime generated by its asymptotic symmetries, the variation of the conserved charges can be calculated. This variation  can be interpreted as an extended first law of thermodynamics that provides an insight into the non-conservation and apparent time-reversal symmetry breaking of this dynamical system. 

This manuscript is organized as follows. In Sect. \ref{AsymSym}, we give a brief description of the general definition of asymptotic symmetries, then proceed by portraying how this definition manifests in general relativity. The remainder of this section is dedicated to specify the near-horizon geometry that we wish to study, and then to derive the generators of the asymptotic symmetries of this geometry. 

In Sect. \ref{EOM}, the equations of motion for this geometry are given where the timelike boundary conditions have been taken into account in the solutions. Sect. \ref{SurfaceCharge} consists of the derivation of the variation in surface charge. This calculation is presented as an analogue of an extended first law of thermodynamics in Sect. \ref{Thermo}. A conclusion and some remarks for possible future work are provided in \ref{Conclusion}.


\section{Asymptotic Symmetries}\label{AsymSym}
Considering a gauge theory on a principal fiber bundle $(P, \mathcal{M},\pi)$, a gauge transformation is defined as an automorphism $\phi : P\rightarrow P$ such that the following diagram commutes

\begin{displaymath}
    \xymatrix{
        P \ar[r]^{\phi} \ar[rd]_{\pi} & P \ar[d]^{\pi} \\
               & M }
\end{displaymath}
A gauge is therefore fixed by choosing a section $S: M\rightarrow P$ of this fiber bundle, and a gauge transformation maps sections to sections. 

One can define a large gauge transformation as a gauge transformation that is not homotopic to the identity map. In terms of physical systems that we deal with, trivial gauge transformations do not change the physical states of a system, whereas large gauge transformations in fact map physical states to different ones. In other words, while a trivial gauge transformation can not affect the physical parameters of the system, a large gauge transformation can influence the physical state.

The gauge transformations in general relativity are diffeomorphisms, and the physical quantities are defined through geometry. A trivial gauge transformation on a manifold $(\mathcal{M},g)$ can be constructed by small diffeomorphisms, generated by exact Killing vector fields of the geometry. On the other hand, large gauge transformations in a diffeomorphism invariant theory such as general relativity, are defined by the quotient space of all possible gauge transformations modulo the trivial gauge transformations of the theory. If the group of trivial gauge transformations is a normal subgroup of all possible transformations, then its quotient space is the group of large gauge transformations. 

To proceed in finding large gauge transformations in a spacetime with geometry $g_{\mu\nu}$, we first fix a coordinate system to discard the gauge freedom related to trivial gauge transformations in GR. Thereafter, to find the quotient space, we need to find all remaining diffeomorphisms that leave the fixed gauge untouched.

\subsection{Gauge fixing}
In the case of the de Sitter horizon defined by the hypersurface $\rho = 0$, we can construct a near-horizon geometry that takes into account non-stationary perturbations of order $\rho$ in the metric field. To define the perturbation parameter, it should be noted that the geometry will not change at the horizon by adding terms of order $\rho$ and higher. Hence, we fix the gauge by defining a near-horizon geometry using the definition in \cite{Chrusciel:2015} defined for any null hypersurface. For the sake of maintaining some level of self-consistency, in the following, we give a brief description of the construction of this coordinate system\footnote{See 
\cite{Chrusciel:2015} for a rigorous description of this construction and further examples}. 

\subsection*{The near-horizon geometry as the gauge}

Let $(\mathcal{M},g)$ be an $4$-dimensional pseudo-Riemannian manifold. By constructing a suitable coordinate system near the horizon, we wish to make the asymptotic symmetries on the horizon manifest.

According to the theorem stated in \cite{Chrusciel:2015}, the geometry near a smooth null hypersurface $\mathcal{H} (\subset \mathcal{M})$ can always be described by the metric $g$ as follows
\begin{equation}
    \begin{split}
        g =& \rho\phi(v,\rho,x)dv^2 + 2dv d\rho + 2\rho h_{A}(v,\rho,x)dvdx^A \\
        &+ g_{AB}(v,\rho,x)dx^{A}dx^{B}
    \end{split}
\end{equation}
where $v$ is the time coordinate, $\rho$ is the radial coordinate, $x=x^A$ ($A=\theta,\phi$) are the angular coordinates on the two-sphere and $\mathcal{H}$ is given by the equation $\rho = 0$.

Specifically, according to the theorem, this geometry is formed by considering a coordinate system $x^A$ on a two-sphere $\mathcal{S}(\subset\mathcal{H})$. Then a coordinate $v$ can be constructed by extending the coordinates $x^A$ to a neighbourhood $\mathcal{H}'(\subset\mathcal{H})$ of $\mathcal{S}$, by demanding $\nabla_{\chi}\chi = 0$ for some vector field $\chi = \chi^\mu\partial_\mu$ with initial value $\chi|_{\mathcal{S}}$, and taking
$\chi(x^A) = \chi^\mu\partial_\mu x^A = 0$. Thus, $\chi |_{\mathcal{H}'}\equiv\partial_v$ is a null generator on $\mathcal{H}'(\subset\mathcal{H})$, which means $g(\chi,\chi)|_{\mathcal{H}'} = g_{vv}|_{\mathcal{H}'} = 0$. Henceforth, it takes the form $g_{vv} \equiv \rho\phi(v,\rho,x)$, where $\phi$ is a function of coordinates and must be finite on the horizon. So far, this gives a coordinate system $(v,x)$ on the horizon. Furthermore, one can deduce $g(\chi,\partial_A)|_{\mathcal{H}'}=g_{vA}|_{\mathcal{H}'} = 0$ from the fact that $\chi |_{\mathcal{H}'}$ is null and thus normal to $\mathcal{H}'(\subset\mathcal{H})$. On this account, $g_{vA} \equiv \rho h_A(v,\rho,x)$, where $h_A$ is also finite at the horizon. 

To extend this coordinate system, we can consider a neighborhood $\mathcal{U}(\subset\mathcal{M})$ of $\mathcal{H}'$  and demand $\nabla_{\Upsilon}\Upsilon=0$ with initial value $\Upsilon|_{\mathcal{H}'}$. Hence, the coordinate system $(v,x)$ on $\mathcal{H}'$ is extended to $(v,\rho,x)$ on $\mathcal{U}$, by solving $\Upsilon(v) = \Upsilon(x^A) = 0$, and defining $\rho$ to be the solution of the equation $\Upsilon(\rho)=1$ with initial value $\rho=0$ on $\mathcal{H}'$, which is the equation defining the null surface $\mathcal{H}$. As a result of this construction, we have $\Upsilon = \partial_{\rho}$, and the $v\rho$ component of the metric is fixed to $g_{v\rho} = 1$. 

For our purpose we work with the following form of the near-horizon geometry to describe the geometry near the de Sitter horizon to first order in $\rho$,
\begin{equation}\label{NHgeo}
    \begin{split}
        ds^2 =& 2\rho\kappa(v,x) dv^2 + 2dvd\rho + 2\rho h_{A}(v,x)dvdx^A \\
        &+ g_{AB}(v,\rho,x)dx^Adx^B,
    \end{split}
\end{equation}
where 
\begin{equation}\label{g_AB}
    g_{AB}(v,\rho,x) = (\rho + l(v))^2 q_{AB}(x),
\end{equation}
and $q_{AB}(x)$ is the metric on the unit two-sphere. We later constrain the variable fields $\kappa$, $h_A$ and $l$ through the equations of motion, to recover pure de Sitter in the limit $v\rightarrow\infty$. 

This construction provides four equations that fix the diffeomorphism gauge freedom appropriate for finding the form of the asymptotic symmetry generators.

\subsection{Asymptotic Symmetry generators}
According to this geometry, the gauge equations are by definition
\begin{equation}
    \begin{split}
        &\delta g_{\rho\rho} = 0, \hspace{6pt} \delta g_{\rho A} = 0, \hspace{6pt} \delta g_{\rho v} = 0\hspace{3pt}
    \end{split}
\end{equation}
with boundary conditions
\begin{equation}
    \begin{split}
        \delta g_{vv} = \mathcal{O}(\rho^2),\hspace{6pt}\delta g_{vA} = \mathcal{O}(\rho^2),\hspace{6pt}\delta g_{AB} = \mathcal{O}(\rho^3).
    \end{split}
\end{equation}

Solving these equations, one can find the form of the generators of the transformations that leave the gauge fixed. They are consequently the generators of the asymptotic symmetry group. This yields, 
\begin{equation}\label{gen}
    \begin{split}
        0 &= \delta g_{\rho\rho} = \mathcal{L}_{\xi}g_{\rho\rho} \\
        &\Longrightarrow \xi^v = f(v,x) \\
        0 &= \delta g_{\rho A} = \mathcal{L}_{\xi}g_{\rho A} \\
        &\Longrightarrow \xi^A = Y^{A}(v,x)- \partial_B f(v,x)\int_{0}^{\rho}g^{AB}d\rho' \\
        0 &= \delta g_{\rho v} = \mathcal{L}_{\xi}g_{\rho v} \\
        &\Longrightarrow \xi^{\rho} = F(v,x) - \rho\partial_{v}f + \partial_Bf\int_{0}^{\rho}g^{AB}g_{vA}d\rho'\hspace{3pt},
    \end{split}
\end{equation}
where $f$, $Y^A$ and $F$ are arbitrary functions of the coordinates $(v,x)$. 

\subsection{Dynamics of Generators}

The variation of the metric field now gives the non-stationary perturbations in the geometry. For the non-zero components of the metric, we thus have 
\begin{equation}
    \begin{split}
        & \delta g_{vv} = 2\rho \delta\kappa \\
        & \delta g_{vA} = \rho \delta h_{A} \\
        & \delta g_{AB} = 2(\rho + l)\delta l q_{AB}\hspace{3pt}.
    \end{split}
\end{equation}
Using the boundary conditions and the field variations through $\delta g_{\mu\nu} = \mathcal{L}_{\xi}g_{\mu\nu}$, leads to 
\begin{equation}
    \begin{split}
        \mathcal{O}(\rho^0)_{\delta g_{vv}} = 2\partial_v F + 2\kappa F = 0.
    \end{split}
\end{equation}
Noting that the asymptotic symmetry generators are taken to be fixed to first order, they do not depend on the dynamical fields \footnote{In \cite{Donnay2016}, they mention this assumption as the \textbf{state independence} of the boundary conditions.}. On this account, we take $F = 0$, and as a result, 
\begin{equation}
    \begin{split}
        \mathcal{O}(\rho^0)_{\delta g_{vA}} = Fh_A + \partial_A F + l^2 q_{AB}\partial_vY^B = 0
    \end{split}
\end{equation}
gives $\partial_v Y^A = 0$. To next order, one has 
\begin{equation}\label{deltaphi}
    \begin{split}
        \mathcal{O}(\rho)_{\delta g_{vv}} &= 2\rho\left[ \partial_v\left(f\kappa\right) + h_A\partial_vY^A -\partial_v^2f \right] \\
        & \Longrightarrow
        \delta\kappa =  \partial_v (f\kappa) -\partial_v^2f.
    \end{split}
\end{equation}
From this equation and the boundary conditions  we can write 
\begin{equation}\label{ST}
    \begin{split}
        f(v,x) = T(x) + Z(v,x),
    \end{split}
\end{equation}
where $T(x)$ is an arbitrary function on the two-sphere at the horizon. This generates a supertranslation-like transformation along the horizon, that causes the transition between different states of the dynamical process. Also, 
\begin{equation}\label{deltag_vA}
    \begin{split}
       \mathcal{O}(\rho)_{\delta g_{vA}} &= \rho \left[ f\partial_v h_{A} + Y^{B}\partial_B h_{A} + h_{B}\partial_{A}Y^{B} - \partial_{A}\partial_v f \right] \\
       &= \rho\delta h_{A}\\
       \Longrightarrow &\hspace{2pt}\delta h_A = f\partial_v h_{A} + Y^{B}\partial_B h_{A} + h_{B}\partial_{A}Y^{B} - \partial_{A}\partial_v f.
    \end{split}
\end{equation}
The last components yield, 
\begin{equation}
    \begin{split}
       \mathcal{O}(\rho^0)_{\delta g_{AB}} &= 2fl\partial_v l q_{AB} + l^2 (D_AY_B + D_BY_A) = 2l\delta l q_{AB} \\ &\Longrightarrow \delta l = f\partial_v l + \frac{1}{2}l\psi,
    \end{split}
\end{equation}
where $\psi \equiv D_AY^A$. 
\section{Equations of Motion}\label{EOM}

In a universe dominated by a cosmological constant, we consider a mass locally centered within its cosmological horizon. The equation of state near the horizon, can be approximated by: 
\begin{equation}\label{eostate}
    \begin{split}
        w=p_{tot}/\epsilon_{tot}&=(-1+\frac{1}{3}\frac{\epsilon_{rad}}{\epsilon_\Lambda})(1+\frac{\epsilon_{rad}}{\epsilon_\Lambda})^{-1}\\
        &\approx-1+\frac{4}{3}\frac{\epsilon_{rad}}{\epsilon_\Lambda}>-1,
    \end{split}
\end{equation}
where $\epsilon $ and $p$ are the energy and pressure densities respectively. In equation (\ref{eostate}) and by the indices ``$\Lambda$" and ``${rad}$", we label the cosmological constant and the radiation emitted out from the mass inside the bulk. Here, the radiation density is not comparable in value with the vacuum energy density, i.e. $\epsilon_{rad}/\epsilon_{\Lambda}\ll 1$. Since this ratio is also decreasing as the constituents of the mass in the bulk gradually settle down in their ground states, one can approximate $T_{\mu\nu}$ at the horizon defined by $\rho = 0$ in (\ref{NHgeo}), as 
\begin{equation}
    T_{\mu\nu}|_{\rho=0}\approx -\epsilon_{\Lambda}\left(1-\frac{1}{3}\frac{\epsilon_{rad}}{\epsilon_\Lambda}\right)g_{\mu\nu}.
\end{equation}
As a consequence, by introducing the function $\lambda(v,x)$ such that
\begin{equation}
    \begin{split}
        \lim\limits_{v\to\infty}\lambda(v,x) = \Lambda_0 = const,
    \end{split}
\end{equation}
the Einstein equations on the horizon can be written as
\begin{equation}
    \begin{split}
        &G_{\mu\nu} + \lambda(v,x)g_{\mu\nu} \approx 0.
    \end{split}
\end{equation}
The function $\lambda(v,x)$ accounts for the effects caused by legitimately small amounts of matter passing through the horizon, radiation, or any other type of perturbation. The constraint on $\lambda(v,x)$ is to ensure pure de Sitter is recovered in timelike infinity\footnote{Although we have considered the fate of the Universe to settle in pure de Sitter at timelike infinity, other scenarios also seem to be possible if one loosens the timelike boundary conditions.}.

Continuing with the Einstein equations, the $vv$ component at the horizon is, 
\begin{equation}
    \begin{split}
        G_{vv}|_{\rho = 0} = -2\kappa\frac{\partial_v l}{l} - 2\frac{\partial_v^2 l}{l} = -\lambda(v,x)g_{vv}|_{\rho = 0} = 0.
    \end{split}
\end{equation}
Since we have taken $l$ to be only a function of $v$, this component of the equations results in $\partial_A\kappa = 0$. For the $\rho v$ component we solve $G_{\rho v} = -\lambda(v,x)g_{\rho v}$ at the horizon to find
\begin{equation}\label{Grhov}
    \begin{split}
        G_{\rho v}|_{\rho = 0} = &-\frac{1}{l^2}(1 - 2 \partial_v l + 2l \kappa - \frac{1}{4}h_Ah^A + \frac{1}{2}q^{AB}\partial_Ah_B \\
        &+ \frac{1}{4} q^{AB}h^C\partial_Cq_{AB}) = -\lambda(v,x). \\
    \end{split}
\end{equation}
As mentioned above, in order to retrieve pure de Sitter as $v\rightarrow\infty$, we wish $\lim\limits_{v\to\infty}\lambda(v,x)= 3/l_0^2$, and $\lim\limits_{v\to\infty} l(v) = l_0 $, where $l_0$ is the cosmological horizon in pure de Sitter. This gives $\lim\limits_{v\to\infty}\kappa = 1/l_0$. We also have,
\begin{equation}
    \begin{split}
        \lim\limits_{v\to\infty}\left[ -h_Ah^A + 2q^{AB}\partial_{A}h_{B} + q^{AB}h^{C}\partial_{C}q_{AB} \right] = 0.
    \end{split}
\end{equation}
The $vA$ components of the EOM  at the horizon reduce to
\begin{equation}\label{GvA}
    \begin{split}
        G_{vA} = \partial_A \kappa - \frac{\partial_vl}{l}h_A - \frac{1}{2}\partial_vh_{A} = -\lambda(v,x)g_{vA}|_{\rho = 0} = 0.
    \end{split}
\end{equation}
Since $\partial_A\kappa = 0$ and $l=l(v)$, this equation gives the $v$ functionality of $h_A(v,x)$, 
\begin{equation}
    \begin{split}
         &\frac{\partial_vl}{l}h_{A} = -\frac{1}{2}\partial_v h_{A} \\
        &\Longrightarrow h_{A}(v,x) = a_{A}(x)b(v) = \frac{a_A(x)}{l(v)^2},
    \end{split}
\end{equation}
where separation of variables is applied to $h_A(v,x)=a_A(x)b(v)$. In turn, the $AB$ components at the horizon are
\begin{equation}\label{GAB}
    \begin{split}
        G_{AB}|_{\rho = 0} = &-\frac{1}{2}h_{A}h_{B}+\frac{3}{4}q^{CD}h_{C}h_{D}q_{AB} \\
        &+ \frac{1}{2}(\partial_{A}h_{B} + \partial_{B}h_{A}) - q^{CD}\partial_{C}h_{D}q_{AB} \\
        &- 2l\kappa q_{AB} - \frac{1}{2}q^{CD}h_{B}\partial_{A}q_{CD} \\
        = &-\lambda(v,x) g_{AB}|_{\rho=0} \\
        = &- \lambda(v,x) l^2 q_{AB}.
    \end{split}
\end{equation}
From relation (\ref{GAB}) we have, 
\begin{equation}\label{AB-lambda}
    \begin{split}
        & -h_{A}h^{A} + \frac{1}{2}q^{AB}(\partial_{A}h_{B}+\partial_{B}h_{A} + q^{CD}h_{B}\partial_{A}q_{CD}) + 4l\kappa \\
        & = 2\lambda(v,x) l^2.
    \end{split}
\end{equation}
As a side note, the reason that the coefficient of $l\kappa$ in (\ref{AB-lambda}) does not match to give $\lim\limits_{v\to\infty} \kappa = 1/l_0$, is that we have not considered the higher order in the $g_{vv}$. If we take $g_{vv} = 2\rho\kappa_1 + \rho^2\kappa_2$, the $G_{AB}$ equations will give 
\begin{equation}
    \begin{split}
         G_{AB} = &-\frac{1}{2}h_{A}h_{B}+\frac{3}{4}q^{CD}h_{C}h_{D}q_{AB} + \frac{1}{2}(\partial_{A}h_{B}+\partial_{B}h_{A}) \\
         &- q^{CD}\partial_{C}h_{D}q_{AB} - 2l\kappa_1 q_{AB} - l^2\kappa_2 q_{AB} \\
         &- \frac{1}{2}q^{CD}h_{B}\partial_{A}q_{CD}, 
    \end{split}
\end{equation}
which results in $\lim\limits_{v\to\infty}\kappa_1 = 1/l_0$ and $ \lim\limits_{v\to\infty}\kappa_2 = 1/l_0^2 $ and we retrieve pure de Sitter as desired. 
\section{Surface Charges}\label{SurfaceCharge}
Having found the generators of the asymptotic symmetries, we can calculate the surface charge using~\cite{Barnich-BMSalgebra}
\begin{equation}
    \begin{split}
        \slashed{\delta} Q_{\xi}[h,g] = &- \frac{1}{16\pi}\int_{S}(d^2x)_{\mu\nu}\sqrt{-g} [D^{\nu}(\xi^{\mu}h) + D_{\sigma}(h^{\mu\sigma}\xi^{\nu}) \\
        &+ D^{\mu}(h^{\nu\sigma}\xi_{\sigma}) + \frac{3}{2}hD^{\mu}\xi^{\nu} \\
        &+ \frac{3}{2}h^{\sigma\mu}D^{\nu}\xi_{\sigma} + \frac{3}{2}h^{\nu\sigma}D_{\sigma}\xi^{\mu} - (\mu \leftrightarrow\nu)],
    \end{split}
\end{equation}
where $ h_{\mu\nu} \equiv \delta g_{\mu\nu} $, $D_{\mu}$ is the covariant derivative of the metric, and
\begin{equation}
    \begin{split}
        (d^2x)_{\mu\nu} = \frac{1}{4}\epsilon_{\mu\nu\sigma\delta}dx^{\sigma}\wedge dx^{\delta}, \hspace{6pt}\epsilon_{v\rho\theta\phi}=1.
    \end{split}
\end{equation}
Note that $\slashed{\delta}$ indicates that the integrability of $\delta Q$ is not determined. 

In our case, the symmetry generators are given by (\ref{gen}), and the corresponding change in the asymptotic charges is given accordingly by
\begin{equation}\label{NEDS-Charge}
    \begin{split}
         \slashed{\delta}Q_{\xi}[h,g] =
        &+\frac{1}{8\pi} \partial_v\int_{\mathcal{H}} d\Omega^2 \sqrt{q} f\delta l^2 + \frac{1}{8\pi} \int_{\mathcal{H}} d\Omega^2 \sqrt{q}f\partial_vl\delta l \\
        &+ \frac{1}{16\pi} \int_{\mathcal{H}} d\Omega^2 \sqrt{q} Y^A\delta (l^2h_A) \\
        & - \frac{1}{4\pi} \int_{\mathcal{H}} d\Omega^2 \sqrt{q}\delta l^2\partial_v f + \frac{1}{8\pi} \int_{\mathcal{H}} d\Omega^2 \sqrt{q}f\kappa \delta l^2. \\
    \end{split}
\end{equation}

A related point to consider is that in the limit $v\to\infty$, $\kappa_0 = 1/l_0$, $\partial_vl=0$, and for $\xi_0 = \partial_v $ ($f=1,Y^A=0$), the expression (\ref{NEDS-Charge}) leads to
\begin{equation}\label{PureCharge}
    \begin{split}
        &\delta Q_{0} = \frac{1}{8\pi}\delta\int_{\mathcal{H}} d\Omega^2 \sqrt{q}(l_0^2\kappa_0)\\
        &\hspace{6pt}\Longrightarrow\hspace{6pt} Q_0 = \frac{l_0}{2}. 
    \end{split}
\end{equation}
Considering $T_0 \equiv (2\pi l_0)^{-1}$ and $S_0=A_{\mathcal{H}}/4 = \pi l_0^2$, $Q_0 = S_0T_0$ can be interpreted as the heat observed by the static patch observer in pure de Sitter ~\cite{HawkingGibbons}. In other words, the static patch observer assigns an apparent temperature to its horizon as a result of its inability to survey the entire spacetime.

\section{Extended First law of Thermodynamics for a dynamical inflating universe}\label{Thermo}

It is well known that there is an analogy between equilibrium thermodynamics and gravity in black holes~\cite{Bekenstein73,Bekenstein1974,Hawking1975,Wald-thermo,Carlip-BHThermo,AshtekarKrasnov,Wald-thermo,Page05,StromingerVafa,Rovelli-QG}. In non-equilibrium thermodynamic cases, the end state of dynamical gravitational systems such as mergers or radiating black holes, can be either a stationary or Minkowski spacetime. In other words, these dynamical systems decay to settle into an analogue of an equilibrium configuration. Undeniably, thermodynamic properties of these systems cannot be described completely through equilibrium thermodynamics. 

For instance, one can refer to~\cite{Freidel-NET} for non-equilibrium thermodynamic properties of an asymptotically flat spacetime containing gravitational radiation. There, gravitational waves have been shown to be an entropy producing mechanism of a viscous dynamical screen.   

In an irreversible thermodynamic process, we have 
\begin{equation}
    \begin{split}
        dS >\frac{dQ}{T},
    \end{split}
\end{equation}
which is due to entropy production in the system. In addition, $dS/dt\neq 0$, which makes any process with a gravitational wave production mechanism, irreversible and non-equilibrium. Here, we show that the relation (\ref{NEDS-Charge}) can be taken as an analogue of an extended first law of thermodynamics, hence indicating the dynamics of our setup to include entropy production. 

Proceeding with the first term in (\ref{NEDS-Charge}), we find that 
\begin{equation}
    \begin{split}
        \frac{1}{2\pi} \partial_v \delta \left[\frac{1}{4} \int_{\mathcal{H}}d\Omega^2\sqrt{q}fl^2 \right] = \frac{1}{2\pi} \partial_v \delta S
    \end{split}
\end{equation}
can be understood as a consequence of the non-conservation of entropy in a dynamical system in an irreversible process. It has a close resemblance to $dS/dt$ which is not zero in non-equilibrium thermodynamics. The existence of $f$ again is due to the infinite dimensional symmetry in the system. 

Taking $\Lambda(v)=3/l^2(v)$ as a state parameter, the second term
\begin{equation}
    \begin{split}
        \frac{1}{8\pi}\int_{\mathcal{H}}d\Omega^2 \sqrt{q}f\partial_vl\delta l &= -\frac{1}{8\pi}\int_{\mathcal{H}}d\Omega^2 \sqrt{q}f \frac{l^3}{6} \partial_vl\delta\Lambda \\
        &=-\int_{\mathcal{H}}d\Omega^2 \sqrt{q}f\frac{l^3}{6}\partial_vl \delta\rho_{\Lambda} \\
        &\equiv \Theta \delta\rho_\Lambda 
    \end{split}
\end{equation}
resembles $\Theta d\Lambda$ in \cite{Sekiwa}\footnote{A similar analysis has been done for anti-de Sitter in~\cite{Kastor}, where the negative cosmological constant is considered as a thermodynamic variable}. Here $\Theta$ is the generalized force conjugate to $\Lambda$, and corresponds dimensionally to a generalized volume. Specifically, $\Theta$ is the generalized time-dependent volume inside the dynamical cosmological horizon up to a factor. Since $\rho_{\Lambda}\equiv  \Lambda / 8\pi $, the term $\Theta\delta\rho_{\Lambda}$ is the contribution of vacuum energy production which is consistent with the increase of vacuum energy in a dynamically inflating universe. As long as $\partial_v l>0$, this term is the contribution of the increase in the volume enclosed by the horizon, due to the time dependent cosmological constant. In the limit $v\rightarrow\infty$, this term vanishes as expected for pure de Sitter spacetime with a constant cosmological constant. 

The term
\begin{equation}
    \begin{split}
        \frac{1}{16\pi}\int_{\mathcal{H}}d\Omega^2 \sqrt{q}Y^A\delta(l^2h_A) \equiv \Omega \delta J 
    \end{split}
\end{equation}
is the contribution of angular momentum to the thermodynamic potential.

The fourth term can be rewritten as
\begin{equation}\label{surfacetension}
    \begin{split}
        -\int_{\mathcal{H}}d\Omega^2 \sqrt{q}\frac{\partial_vf}{4\pi f} f\delta l^2 \equiv \sigma \delta A
    \end{split}
\end{equation}
and can be thought of as an analogue of the contribution of surface tension when considering a screen or bubble. It has been shown that this contribution is $\sigma\delta A$, where $\sigma$ is the surface tension and $A$ is the area of the horizon~\cite{Freidel-NET,Callaway}. In equation (\ref{surfacetension}), the surface tension is defined as $\sigma_{\mathcal{H}}\equiv -\partial_v f / 4\pi f$, and can be understood as a negative pressure. This term acts as the work done by a repulsive force, expanding the horizon. Our result is different from that of \cite{CHEN2017115} since we have assumed $T_{\mu\nu} = 0$ near the horizon, while their definition of surface tension is proportional to $T^r_r|_{r_c}$, where $r_c$ is the radius of the cosmological horizon in their coordinates. Accordingly, that contribution is not included in our case. The reason we have not combined this term with (\ref{SdA}) is based on the assumption of taking the surface gravity to be only a function of $v$. 

Finally, in the term 
\begin{equation}\label{SdA}
    \begin{split}
        \frac{\kappa}{2\pi}\frac{1}{4} \int_{\mathcal{H}}d\Omega^2\sqrt{q}f \delta l^2 \equiv T\delta S,
    \end{split}
\end{equation}
we interpret $\kappa/2\pi$ as the temperature of the dynamical inflating spacetime and the expression $\int d\Omega^2\sqrt{q}f\delta l^2$ can be taken as the change in the area. Consequently, this term can be interpreted as the equivalent of $T\delta S$. The coefficient $f(v,x)$ is defined by (\ref{ST}), and this can be a proposal for the dynamical entropy and dynamical surface gravity. Note that $\kappa$ is restricted by the equations of motion and is not an arbitrary function. It is also worth pointing out, this term contributes to non-integrability due to the fact that $\delta\kappa \neq 0$. This issue will become crucial in the study of charge algebras, which will be studied in future work and is not a subject in this work.

In a gravitational system such as a merger, or a ringing black hole, entropy production is implemented by gravitational waves. Here we have found, the stretching of spacetime itself contributes to entropy production, when dealing with an otherwise empty spacetime with positive curvature due to a positive cosmological constant. 

The extended first law of thermodynamics for such a system is, therefore,
\begin{equation}\label{Extended-2L}
    \begin{split}
        \delta Q = \frac{1}{2\pi}\partial_v \delta S + \Theta \delta\rho_\Lambda + \Omega \delta J + \sigma\delta A + T\delta S .
    \end{split}
\end{equation}

As a result, what is found from (\ref{NEDS-Charge}) and consequently (\ref{Extended-2L}), is that the dynamical spacetime of our setup is an analogue of a non-equilibrium process.  

\section{Conclusion}\label{Conclusion}
In this paper, we propose a process described by the equations of motion for the near-horizon geometry of a universe with a dynamical cosmological constant. We consider this system to approach pure de Sitter at timelike infinity, and find that such a system enjoys a larger symmetry than the usual $SO(4,1)$ symmetry of pure de Sitter given by (\ref{gen}). This larger symmetry consists of supertranslation-like generators on the horizon. The supertranslations generate the dynamical transition between different states of an irreversible process. By calculating the change in the charges associated to these symmetry generators, we find an extended version of the first law of thermodynamics, expressed in (\ref{Extended-2L}). This displays the non-conservation of the charges in a non-equilibrium process. The apparently non-conservation of the charges is due to the dynamical aspect we have assumed and the incapability of the observer to examine the degrees of freedom in the entire spacetime beyond its horizon.

\section*{Acknowledgement}
Part of this project was done when F.M. was a fellow at \textit{Center for the Fundamental Laws of Nature, Harvard University} in 2018. F.M. would like to thank Andrew Strominger for providing this incredible opportunity.

\nocite{Ferreira16}
\bibliography{main}

\end{document}